\documentclass[a4paper]{article}

\usepackage{graphicx}
\usepackage{onecolpceurws}

\usepackage{amsfonts}
\usepackage{dsfont}

\usepackage{mathtools}
\usepackage{amssymb}
\usepackage{amsmath}
\usepackage{amsthm}
\usepackage{epstopdf}
\usepackage{colortbl}

\usepackage{enumitem} 
\usepackage{caption}
\usepackage[table,xcdraw]{xcolor}

\usepackage{float}
\usepackage{subcaption}
\usepackage{epsfig}
\usepackage{multirow}
\usepackage[all]{xy}
\usepackage{mdframed}
\usepackage{multirow}
\usepackage{url}
\usepackage{lipsum}

\title{Formalization of Transform Methods in Higher-order Logic: A Survey}

\author{
Muhammad Ahmed and Adnan Rashid
}

\institution{School of Electrical Engineering and Computer Science (SEECS) \\
                National University of Sciences and Technology (NUST) \\
                Islamabad, Pakistan \\ \{mahmed.mscs19seecs,adnan.rashid\}@seecs.nust.edu.pk}

\begin{document}
\maketitle

\begin{abstract}
Most of the engineering and physical systems are generally characterized by differential and difference equations based on their continuous-time and discrete-time dynamics, respectively. Moreover, these dynamical models are analyzed using transform methods to prove various properties of these systems, such as, transfer function, frequency response and stability, and to find out solutions of the differential/difference equations.
The conventional techniques for performing the transform methods based analysis have been unable to provide an accurate analysis of these systems. Therefore, higher-order-logic theorem proving, a formal method, has been used for accurately analyzing systems based on transform methods. In this paper, we survey developments for transform methods based analysis in various higher-order-logic theorem provers and overview the corresponding real world case studies from the avionics, medicine and transportation domains that have been analyzed based on these developments.
\end{abstract}
\vskip 32pt

\section{Introduction}\label{SEC:Intro}

\begin{figure}[!ht]
  \centering
  \includegraphics[width=\linewidth]{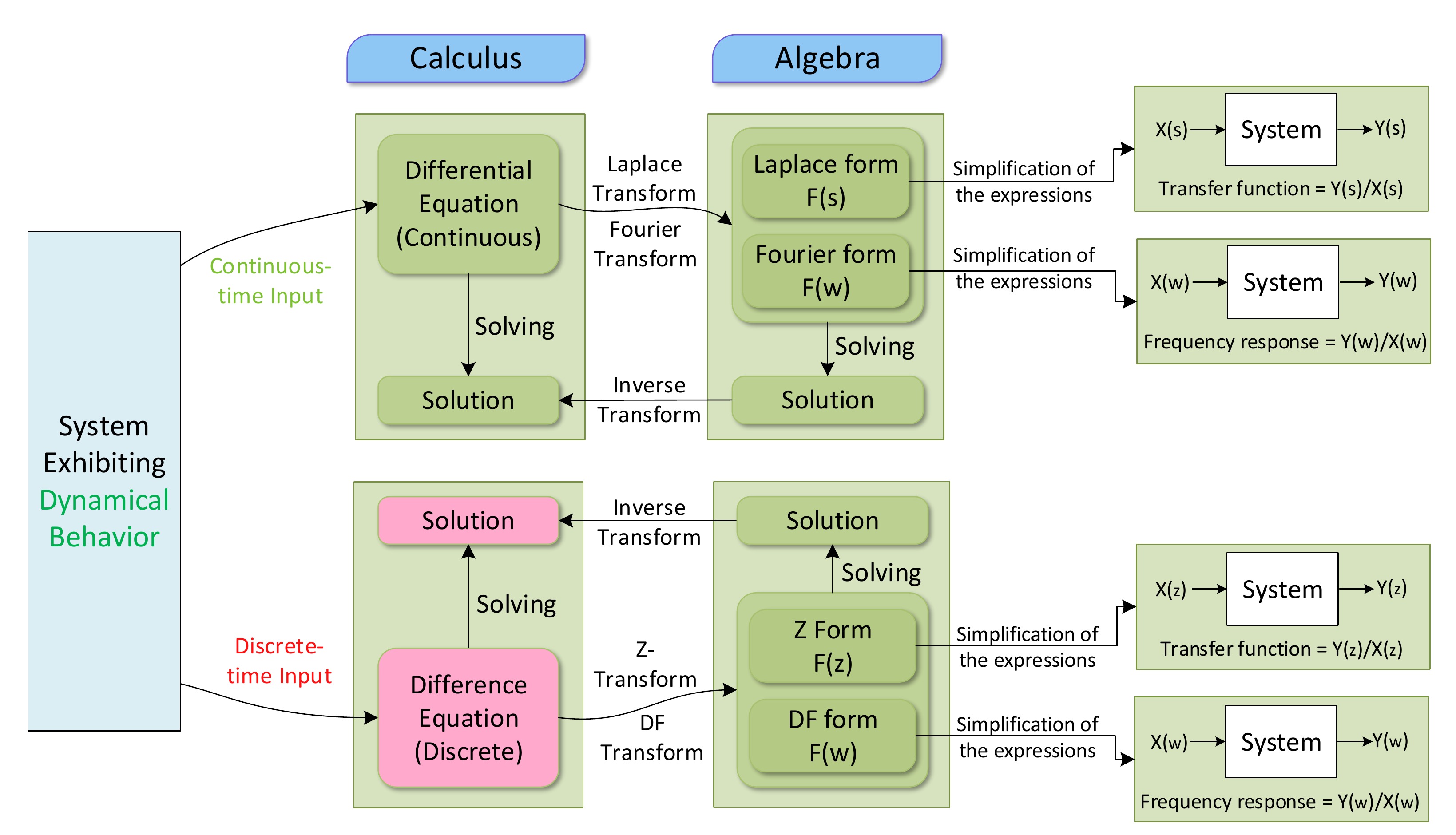}
  \caption{Transform Methods}
  \label{fig:transform_methods}
\end{figure}

The engineering and physical systems exhibiting dynamical behaviours are generally modeled using differential~\cite{yang2013mathematical} and difference~\cite{kelley2001difference} equations based on their corresponding dynamics that can be continuous-time  and discrete-time, respectively. To analyze these models capturing the dynamics of such systems, transform methods, such as, the Laplace transform~\cite{beerends2003fourier}, the Fourier transform~\cite{bracewell1986fourier}, the Discrete Fourier Transform (DFT)~\cite{sundararajan2001discrete} and the $z$-Transform~\cite{oppenheim2001discrete}, have been widely used. These transform methods facilitate finding out solutions of these differential and difference equations based models and analyzing their various important properties, such as, transfer function, frequency response and stability, as shown in Figure~\ref{fig:transform_methods}.

The continuous-time dynamical behaviour of a system is generally modeled using differential equations. These differential equations-based models are not easier to analyze in time-domain, especially, for the case of larger systems ($n$-order differential equations for larger $n$). Transform methods, which include the Laplace and the Fourier transforms have been widely used to analyze these differential equations. These methods involve a transformation of the time-domain model to its corresponding frequency domain representation ($s$-domain for the Laplace and $\omega$-domain for the Fourier transform), which converts the differential equations involving integrals and differentials to their corresponding algebraic equations having multiplication and division operators, as shown in Figure~\ref{fig:transform_methods}. These equations are further solved to analyze various properties of the system, such as, transfer function and frequency response, and to obtain solutions in frequency domain~\cite{beerends2003fourier,bracewell1986fourier}. Finally, the inverse transforms (the inverse Laplace and the inverse Fourier) are applied to obtain the time-domain solutions of the differential equations-based models.

Similarly, the dynamics of a discrete-time system are captured using difference equations. These difference equations based models are analyzed using the discrete-time transform methods, which include the Discrete Fourier Transform (DFT) and the $z$-transform. These transform methods perform a conversion of the discrete time model to its corresponding frequency domain representation ($z$-domain for the $z$-transform and $\omega$ for the DFT), as shown in Figure~\ref{fig:transform_methods}. These representations are further solved to analyze various properties of the given system, such as, transfer function and frequency response, and to obtain solutions in frequency domain~\cite{sundararajan2001discrete,oppenheim2001discrete}. Finally, the inverse transforms are applied to obtain the time-domain solutions of the difference equations based models.

Conventionally, the transform methods based analysis has been performed using paper-and-pencil based proofs and computer based symbolic and numerical techniques. However, these methods suffer from their inherent limitations and thus compromise the accuracy of the analysis~\cite{duran2013misfortunes}. Formal methods~\cite{hasan2015formal}, in particular, higher-order-logic theorem proving~\cite{harrison2009handbook} can overcome the limitations of the conventional methods and thus provides an accurate transform methods based analysis of the systems. It has been widely used for the transform methods based analysis of the engineering and physical systems. These transform methods have been formalized using various higher-order-logic theorem provers, such as, HOL Light~\cite{harrison1996hol}, HOL4~\cite{slind2008brief}, Coq~\cite{bertot2013interactive}
and Isabelle~\cite{paulson1994isabelle}.
Moreover, these formalizations have been used for analyzing various safety-critical systems, such as, linear analog circuits, automobile suspension systems, a drug therapy model, unmanned aerial vehicles, synthetic biological circuits, power converters and digital filters. In this paper, we survey these contributions regarding formalization of transform methods that have been done using higher-order-logic theorem proving.

The rest of the paper is organized as follows: Section~\ref{SUBSEC:continuous_transform_methods} provides the developments of the Laplace and the Fourier transforms in various theorem provers and their associated analysis of the continuous-time systems. We provide the formalization of the DFT and the $z$-transform and the associated analysis of the discrete-time systems in Section~\ref{SUBSEC:discrete_transform_methods}. Section~\ref{SEC:discussion} presents a discussion about the features and availability of the transform methods based analysis and its comparison in different theorem provers.
Finally, Section~\ref{SEC:conclusion} concludes the paper.

\section{Transform Methods for Analyzing Continuous-time Systems} \label{SUBSEC:continuous_transform_methods}

The Laplace and the Fourier transforms have been formalized using various higher-order-logic theorem provers. Moreover, these formalizations have been used for formally analyzing many engineering and physical systems. Taqdees et al.~\cite{taqdees2013formalization} formalized the Laplace transform using multivariate calculus theories of HOL Light. This formalization mainly includes the formal definition of the Laplace transform and the formal verification of its various properties, such as, linearity, frequency shifting, first-order differentiation, higher-order differentiation and integration in time-domain. Moreover, the authors used their formalization of the Laplace transform for formally verifying the transfer function of a Linear Transfer Converter (LTC) circuit. Next, the authors extended their framework by providing a support to formally reason about the linear analog circuits, such as first-order and second-order Sallen-Key low-pass filters by formalizing the system governing laws such as Kirchhoff’s Current Law (KCL) and Kirchhoff’s Voltage Law (KVL) using HOL Light~\cite{taqdees2017formally}. Later, Rashid et al.~\cite{rashid2017formal} proposed a new formalization of the Laplace transform based on the notion of sets and verified some more properties of the Laplace transform, such as, time shifting, time scaling, the Laplace transform a $n$-order differential equation and uniqueness~\cite{rashid2018formalization}. The authors also formally verified the Laplace transform of some commonly used functions, such as, exponential function, sine and cosine functions. Finally, they used their proposed formalization of the Laplace transform for analyzing the control system of an Unmanned Free-swimming Submersible (UFSS) vehicle~\cite{rashid2017formal} and 4-$\pi$ soft error cross talk model~\cite{rashid2018formalization}. Similarly, the formalization of the Laplace transform has been used for formally analyzing the unmanned aerial vehicles~\cite{abed2020formal} and synthetic biological circuits~\cite{abed20201formal,rashid2020formal}.

Wang et al.~\cite{wang2017formalization} provided a formalization of the Laplace transform using the Coq theorem prover. The authors formally defined the Laplace transform and formally verified a few of its classical properties, such as, linearity, frequency shifting and differentiation in time domain. Moreover, they applied their proposed formalization for formally verifying a flight control system using Coq. Similarly, Immler~\cite{immler2021laplace} formalized the Laplace transform using the Isabelle theorem prover. The authors mainly verified some of the classical properties of the Laplace transform, such as, linearity, frequency shifting, uniqueness, differentiation and integration in time-domain. Gang et al.~\cite{gang2014formalization} used HOL4 for the formalization of the Laplace transform. The authors formally modeled the Laplace transform and verified some of its classical properties, such as, linearity, frequency shifting, differentiation and integration in time domain. Next, they used their proposed formalization for formally verifying the transfer function of a motor.

\begin{table}[!ht]
\caption{Continuous-time Transform Methods in Various Theorem Provers}
\label{tab:my-table1}
\centering
\begin{tabular}{|c|c|c|}
\hline
\rowcolor[HTML]{FFFFC7}
Transform Methods                   & Theorem Prover & Formalization Details/ Verified Properties                                                                                                                     \\ \hline
                                    & Coq            &
                                    Linearity, frequency shifting, differential~\cite{wang2017formalization}                                                                                                               \\ \cline{2-3}
                                    & HOL Light           & \begin{tabular}[c]{@{}c@{}}
                                    Linearity, frequency shifting, integration, time shifting, \\
                                    time scaling, first-order and higher-order \\
                                    differentiations in time domain~\cite{taqdees2013formalization,rashid2018formalization}
                                    \end{tabular}                 \\ \cline{2-3}
                                    & Isabelle           & \begin{tabular}[c]{@{}c@{}}
                                    Linearity, frequency shifting, integration, \\
                                    first-order and higher-order differentiations \\
                                    in time domain, integration~\cite{immler2021laplace}
                                    \end{tabular}                 \\ \cline{2-3}
\multirow{-8}{*}{Laplace Transform} & HOL4           & \begin{tabular}[c]{@{}c@{}}
                                    Linearity, frequency shifting, differentiation and  \\
                                    integration in time domain~\cite{gang2014formalization}  \end{tabular}               \\ \hline
                                    & HOL Light           & \begin{tabular}[c]{@{}c@{}}
                                    Linearity, frequency shifting, modulation, time reversal, \\
                                    first-order and higher-order differentiations \\
                                    in time domain~\cite{rashid2019formal}
                                    \end{tabular} \\ \cline{2-3}
\multirow{-4}{*}{Fourier Transform} & HOL4           & \begin{tabular}[c]{@{}c@{}}
                                    Linearity, time reversal, frequency shifting, differentiation, \\
                                    and integration in time domain~\cite{guan2020formalization}
                                    \end{tabular}                                              \\ \hline
\end{tabular}
\end{table}

Rashid et al.~\cite{rashid2016formalization} formalized the Fourier transform using the HOL Light theorem prover. The authors provided a formal definition of the Fourier transform and formally verified its various classical properties, such as, linearity, frequency shifting, modulation, time reversal, first-order and higher-order differentiations in time-domain. Furthermore, they formally verified the Fourier transform of some commonly used functions, such as, exponential, sine and cosine functions. Moreover, they used their proposed formalization for formally analyzing an  Automobile  Suspension System (ASS), an audio equalizer, a drug therapy model and a MEMs accelerometer~\cite{rashid2019formal}. Similarly, Guan et al.~\cite{guan2020formalization} formalized the Fourier transform using HOL4. The authors provided a formal definition of the Fourier transform and formally verified its various properties, such as, linearity, time reversal, frequency shifting, differentiation and integration in time domain. Moreover, they used their proposed formalization for formally verifying the frequency response of a RLC circuit. A summary of the formalizations of the Laplace and the Fourier transforms in various theorem provers and the systems that have been formally analyzed using these transform methods can be found in Tables~\ref{tab:my-table1} and~\ref{tab:my-table2}

\begin{table}[!ht]
\caption{Analysis of Continuous-time Systems using Transform Methods}
\label{tab:my-table2}
\centering
\begin{tabular}{|c|c|c|}
\hline
\rowcolor[HTML]{FFFFC7}
Transform Methods                   & Theorem Provers & Applications                                                                                            \\ \hline
                                    & Coq             & Flight control system~\cite{wang2017formalization}                                                                                                               \\ \cline{2-3}
                                    & HOL Light       & \begin{tabular}[c]{@{}c@{}}
                                    LTC~\cite{taqdees2013formalization},
                                    UFSS vehicle~\cite{rashid2017formal}, \\
                                    4-$\pi$ soft error cross talk model~\cite{rashid2018formalization}, \\
                                    Sallen-Key low-pass filter~\cite{taqdees2017formally},  \\
                                    unmanned aerial vehicles~\cite{abed2020formal},  \\
                                    synthetic biological circuits~\cite{abed20201formal,rashid2020formal}  \end{tabular}                                            \\ \cline{2-3}
                                    & Isabelle             & No application                                                                                                                 \\ \cline{2-3}
\multirow{-8}{*}{Laplace Transform} & HOL4        & Motor~\cite{gang2014formalization}                                                                                                                        \\ \hline
                                    & HOL Light       & \begin{tabular}[c]{@{}c@{}}
                                     Automobile suspension system~\cite{rashid2016formalization}, \\
                                     a drug therapy model~\cite{rashid2019formal}, \\
                                     an audio equalizer~\cite{rashid2019formal}, \\
                                     MEMs accelerometer~\cite{rashid2019formal}
                                     \end{tabular} \\ \cline{2-3}
\multirow{-5}{*}{Fourier Transform} & HOL4            & RLC circuit~\cite{guan2020formalization}                                                                                                             \\ \hline
\end{tabular}
\end{table}

\section{Transform Methods for Analyzing Discrete-time Systems} \label{SUBSEC:discrete_transform_methods}

The DFT and $z$-Transform have been formalized using various higher-order-logic theorem provers. Moreover, these formalizations have been used for formally analyzing many discrete-time systems. Siddique et al.~\cite{siddique2014formalization} formalized $z$-transform using the HOL Light theorem prover. The authors provided a formal definition of the $z$-transform and formally verified its various properties, such as, linearity, time shifting and scaling in $z$-domain. Moreover, they used their proposed formalization for the formal analysis of Infinite Impulse Response (IIR) Digital Signal Processing (DSP) filter. Later, the authors extended their proposed framework by providing the formal verification of some more properties, such as, time scaling, complex conjugate and a formal support for the inverse z-transform and used it for formally analyzing a switched-capacitor interleaved DC-DC voltage doubler~\cite{siddique2018formal}.

\begin{table}[!ht]
\caption{Discrete-time Transform Methods in Various Theorem Provers}
\label{tab:my-table3}
\centering
\begin{tabular}{|c|c|c|}
\hline
\rowcolor[HTML]{FFFFC7}
Transform Methods                   & Theorem Prover & Formalization Details/ Verified Properties                                                                                                                     \\ \hline
    DFT                             & HOL4           & \begin{tabular}[c]{@{}c@{}}
                                                       Implicit periodicity, linearity, symmetry, frequency shifting, \\
                                                       time shifting, convolution~\cite{shi2015formalization}
                                                    \end{tabular}                                                                                                    \\ \hline
                                    & HOL4           & \begin{tabular}[c]{@{}c@{}}
                                                     FFT, inverse FFT~\cite{akbarpour2004verification}
                                                     \end{tabular} \\ \cline{2-3}
\multirow{-2}{*}{FFT}               &  Coq          & \begin{tabular}[c]{@{}c@{}}
                                            FFT~\cite{capretta2001certifying}
                                          \end{tabular}    \\ \hline
           $z$-Transform            & HOL Light           & \begin{tabular}[c]{@{}c@{}}
                                                       linearity, time shifting and scaling in $z$-domain, \\
                                                       time scaling, complex conjugation, inverse $z$-transform~\cite{siddique2018formal}
                                                       \end{tabular}                                              \\ \hline
\end{tabular}
\end{table}

Shi et al.~\cite{shi2015formalization} proposed a formalization of DFT using the HOL4 theorem prover. The authors presented a formal definition of the DFT and formally verified its various properties, such as, implicit periodicity, linearity, symmetry, frequency shifting, time sifting and convolution. Moreover, the authors used their proposed formalization for formally verifying Fast Fourier Transform (FFT) and cosine frequency shifting. Capretta et al.~\cite{capretta2001certifying} formally verified the FFT using the Coq theorem prover. Similarly, Akbarpour et al.~\cite{akbarpour2004verification} provided a formal specification and verification of the FFT at different abstraction levels using the HOL4 theorem prover. A summary of the formalization of the transform methods for discrete-time systems in various theorem provers and their associated applications can be found in Tables~\ref{tab:my-table3} and~\ref{tab:my-table4}.

\begin{table}[!ht]
\caption{Analysis of Discrete-time Systems using Transform Methods}
\label{tab:my-table4}
\centering
\begin{tabular}{|c|c|c|}
\hline
\rowcolor[HTML]{FFFFC7}
Transform methods     & Theorem Provers & Formalization Details / Verified Properties                                      \\ \hline
DFT                   & HOL4            & FFT, cosine frequency shifting~\cite{shi2015formalization}                                               \\ \hline
                      & HOL4            & FFT at different levels of abstraction and inverse FFT~\cite{akbarpour2004verification}  \\ \cline{2-3}
\multirow{-2}{*}{FFT} & Coq             & FFT~\cite{capretta2001certifying}                                                                            \\ \hline
Z-Transform           & HOL Light       & \begin{tabular}[l]{@{}l@{}}
                                          IIR DSP filter~\cite{siddique2014formalization}, switched-capacitor \\ interleaved DC-DC voltage doubler~\cite{siddique2018formal}  \end{tabular}  \\ \hline
\end{tabular}
\end{table}

\section{Theorem Proving Support for Transform Methods based Analysis} \label{SEC:discussion}

\begin{figure}[!ht]
  \centering
  \includegraphics[width=0.8\linewidth]{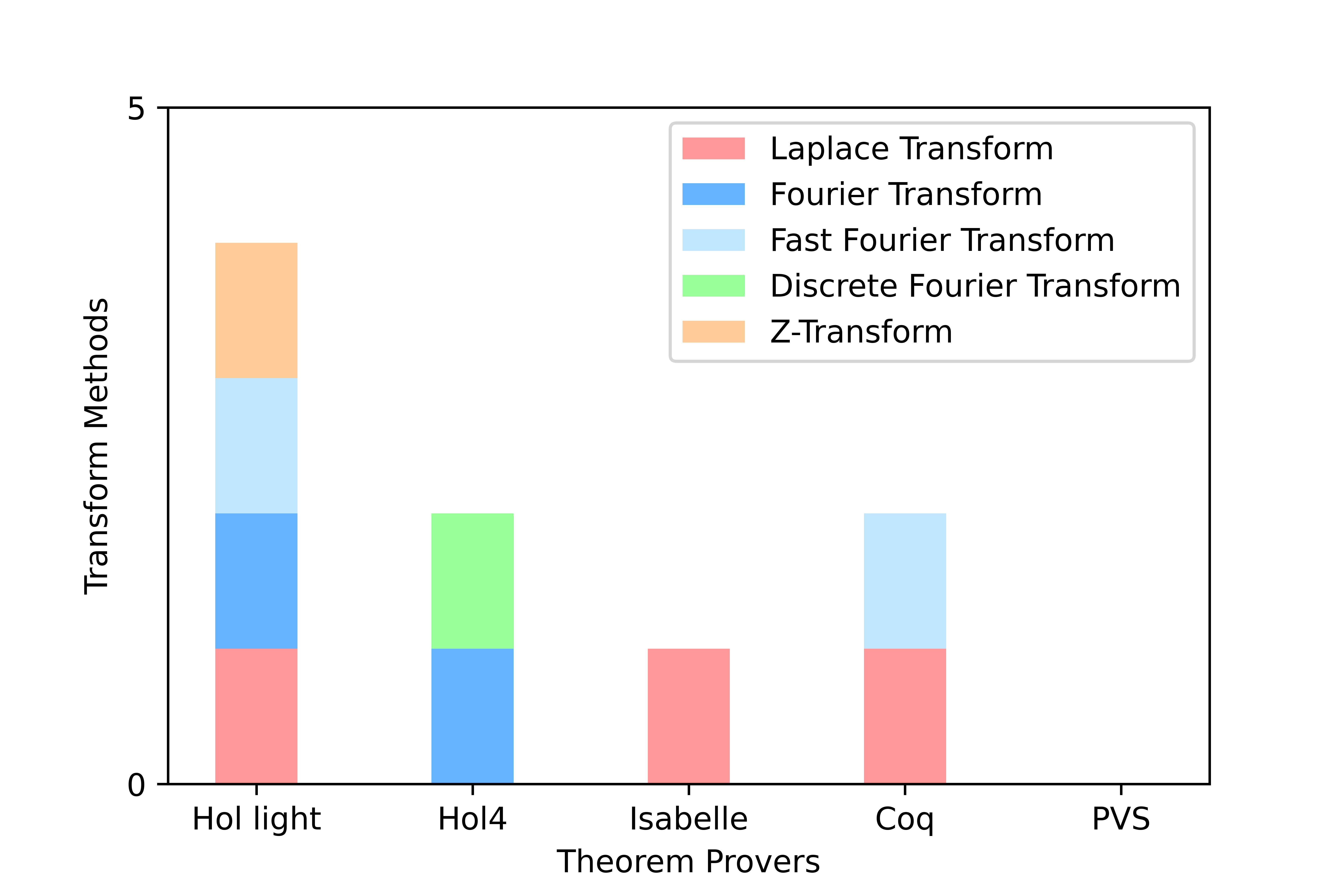}
  \caption{Support for Transform Methods in Higher-order-logic Theorem Provers}
  \label{fig:transform_methods_hol_tp}
\end{figure}

Figure~\ref{fig:transform_methods_hol_tp} depicts the formal libraries of transform methods that are available in various higher-order-logic theorem provers for performing the analysis of the engineering and physical systems. For example, the Laplace transform is available in most of the theorem provers. Moreover, its formal library in the HOL Light theorem prover is quite dense and has been frequently used for analyzing various safety-critical systems as given in Table~\ref{tab:my-table2}. Similarly, 
the $z$-transform is only available in HOL Light. Similarly, HOL4 contains both the continuous-time and discrete-time Fourier transforms. Moreover, no transform methods is available in the PVS theorem prover.

\section{Conclusion}\label{SEC:conclusion}

Transform methods are widely used for analyzing the engineering and physical systems exhibiting dynamical behaviours. Due to the safety-critical nature of these system, their accurate analysis is of utmost importance.  This paper surveys the transform methods that have been formalized using different theorem provers by highlighting various safety-critical systems that have been formally analyzed based on transform methods. In this regard, only HOL Light theorem prover contains most number of transform methods, such as, the Laplace and the Fourier transforms, and the $z$-transform. Similarly, PVS contains no transform methods library and we need to develop these libraries that can be used for performing transform methods based analysis using PVS.

\bibliographystyle{alpha}
\bibliography{biblio}

\end{document}